\documentclass[acmsmall]{acmart}
\usepackage{booktabs} 

\usepackage{tikz}
\usepackage{float}
\usetikzlibrary{shapes,arrows}
\usepackage{graphicx}
\graphicspath{{img/}}
\usepackage{rotating}
\usepackage{localmacros}
\graphicspath{ {./img/} }
\usepackage{xstring}
\usepackage{amsmath}
\usepackage{amsthm}
\usepackage{url}
\usepackage{multirow}
\usepackage{cleveref}
\usepackage[utf8]{inputenc}
\usepackage{pmboxdraw}
\usepackage{subcaption}
\usepackage{pdflscape}
\usepackage{afterpage}
\usepackage{listings}
\makeatletter 
\def\arcr{\@arraycr}
\makeatother

\begin{document}
\title{Preprocessing Source Code Comments for Linguistic Models}


\author{Sergey Matskevich}
\affiliation{%
  \institution{Drexel University}
  \city{Philadelphia}
  \state{Pennsylvania}
  \country{USA}
}
\email{sm3372@drexel.edu}

\author{Colin S. Gordon}
\orcid{0000-0002-9012-4490}
\affiliation{%
  \institution{Drexel University}
  \city{Philadelphia}
  \state{Pennsylvania}
  \country{USA}
}
\email{csgordon@drexel.edu}

\begin{abstract}
Comments are an important part of the source code and are a primary source of documentation. This has driven interest in using large bodies of comments to train or evaluate tools that consume or produce them --- such as generating oracles or even code from comments, or automatically generating code summaries.
Most of this work makes strong assumptions about the structure and quality of comments, such as assuming they consist mostly of proper English sentences. However, we know little about the actual quality of existing comments for these use cases.
Comments often contain unique structures and elements that are not seen in other types of text, and filtering or extracting information from them requires some extra care.
This paper explores the contents and quality of Python comments drawn from 840 most popular open source projects from GitHub and 8422 projects from SriLab dataset, and the impact of na\"ive vs. in-depth filtering can have on the use of existing comments for training and evaluation of systems that generate comments. 
\end{abstract}
\maketitle

\newtheorem{rquestion}{Research Question}
\newcommand{\resq}[1]{
    \IfEqCase{#1}{%
    {1}{What categories of comments exist that may be worth excluding from training data for general-purpose linguistic processing for English?}%
    {2}{Can comment categories worth excluding be automatically and effectively detected and excluded using deterministic techniques?}%
    {3}{How prevalent are the major categories of comments?}%
    {4}{Are there comments which are both high quality but also not useful for direct linguistic processing?}%
    {5}{What is the impact of different levels of filtering on the quality of trained linguistic models for comment completion?}%
    }[\PackageError{resq}{Undefined research question}]
}

\section{Introduction}

Comments play a major role in the software development process. They are often considered to be de-facto documentation which is the closest to the source and play an essential role in understanding~\cite{docStudy} and maintaining~\cite{Aggarwal:maint} software.  Due to the explosion of machine learning and the parallel increase in the number of public online code repositories, much research is now focused on using existing source code for mining, and for training or evaluating automation tools. This work spans such topics as generating partial oracles~\cite{genExceptions}, generating specifications~\cite{comToSpec,Zhai:Com2Spec}, predicting or generating comments~\cite{predictComments,nl4se18,neuralModelGen}, and evaluating the quality of the source or comments~\cite{javaDocQuality,qualityOfComments}.
All of these lines of work rely on source code comments for training and/or evaluation, and many use source code as well.
However, while the fact that code is designed specifically for machine-processing (e.g., by type-checkers, compilers, and unit tests) imposes a kind of lower-bound on code quality, no such baseline is imposed for source code comments.

Many systems that automatically process comments make assumptions about the structure and the contents of the comments. They generally assume that comments are written in mostly plain text (usually in English as opposed to another natural language), optionally assuming light use of highly-standardized structured comments (e.g., Javadoc) or that the full general summary of the whole comment is in the first sentence of the comment. 
However, this is simply assumed, not validated, and most evaluations proceed with small carefully-curated codebases for which these assumptions hold.
If these assumptions do not hold widely, then attempting to train similar tools on different datasets can produce models that have vastly different performance. 

Overall, there is limited comprehensive research into comment quality: the quality of the language in comments, what information is contained in the comments, types of comments, and so on. The comment quality evaluation work that does exist focuses on detailed comment type taxonomies~\cite{commentClassification}, or a holistic way of evaluating whether a comment is good \emph{for software developers on the relevant project}~\cite{javaDocQuality,qualityOfComments}; they do not consider the suitability of comment data for automated processing. There is very recent work~\cite{chen2021my} showing that automatic comment generation is sensitive to comment type, highlighting one way in which unvalidated assumptions of one body of work~\cite{leclair2020improved,LeClairNeuralModel,iyer2016summarizing,movshovitz2013natural} likely skew results.

Most published work consuming comments as training data either performs only basic text sanitization and filtering, or works only with smaller numbers of manually-curated high quality exemplars of comments. Common data sanitization for NLP tasks is simply removing the punctuation, making all words lower case and removing phrases/sentences that are too short. However, this filtering assumes what remains afterwards is linguistic in nature, which we demonstrate is often not true. In this work we set out to test the impact that the data will have on the accuracy of text generation linguistic models with different stages of data sanitization.

This paper analyzes the source comments from the two datasets. One consists of the top 840 open source Python projects on GitHub\footnote{https://github.com/} (208661  files, 46710511 lines of code), and the other comes from the existing SRILab Py150\footnote{https://www.sri.inf.ethz.ch/py150} dataset with 8422 projects containing 149970 files and 24540627 lines of code. We provide a new taxonomy of comments with respect to relevance for \emph{building linguistic models of source code comments}, and investigate the impact different types of comments have on machine learning algorithms that make use of natural language models. Our focus in this paper is not the evaluation of the models as a whole, but a specific sub-task required for building any model: the sanitization of comments with the goal of producing suitable training (and evaluation) data for machine learning based systems.

In this paper we build two language models that we use for generation of comments and evaluate the quality of the generated sentences based on different levels of sanitization of input. The contributions of this paper are:
\begin{itemize}
    \item We explore what types of comments are in the source code (e.g. copyright, special file headers, etc) with respect to suitability for building linguistic models, and provide a taxonomy of relevant types.
	\item We analyze the frequency of these comment types in two large Python corpora.
	\item We analyze the impact of different levels of care in filtering out non-linguistic comments have on machine learning algorithms trained on source code comments.
	\item We highlight discrepancies between developer-focused comment taxonomies and tool-focused taxonomies, discussing the usefulness of the comment to a human vs.\ a machine learning algorithm, finding they are often in opposition.
\end{itemize}
In addition, this paper is to the best of our knowledge, the first analysis of Python code comments (as opposed to prior work, which studies primarily Java and C++), finding that there is a higher incidence of unambiguously useless comments than prior studies have found.

\section{Background and Related Work}
Since machine learning models learn on large amounts of data, one major source of model bias is duplication of data. \citet{dupComments} explored the influence of code duplication on machine learning models trained on large datasets. He points out that raw training data often has many duplicates, which can result in duplication within and across the training and testing datasets, which both contradicts a common assumption in mining work, and can heavily skew results by biasing training towards performing well on duplicated code. He analyzed multiple corpora of different languages and found that the number of duplicates within one dataset can be as much as 25\%, affecting performance metrics by up to 50\%.
Although his work is not directly concerned with comments, the same problem exists for systems trained on comment data. As the dataset contains duplicate source files, all comments in those files will also be duplicated and thus impact training models.  In our study we found that comments from open source repositories contain significant amount of duplication.

\citet{comToSpec} and \citet{genExceptions} use comments to create tools and applications for the automated generation of the specifications and tests. They parse comments, relying on the use of Javadoc formatting  (particularly the \lstinline[language=Java]|@param| tag) to extract oracle information for generating tests. The Javadoc assumption is useful, but also a significant limitation: if the assumed Javadoc formatting was not used, the tools behave as if the information was missing --- even if present in prose --- and will not generate specifications or tests.
The assumption of the structure of comments is also an issue in the context of using the comment for machine learning. We find that comments can have unexpected elements in them, such as text in a multiple natural languages, Markdown, \LaTeX, and even different forms of the source code (\figref{sageMath}).
\citet{pandita:specGen} generate specifications for an API using the API's documentation, though without assuming specific language idioms or comment formatting (i.e., Javadoc).
They expect, however, to find similar information such as method description, argument description, return information, exceptions, and general remarks.

\citet{LeClair:CodeGNN} produced a tool CodeGNN\footnote{\url{https://github.com/acleclair/ICPC2020\_GNN}} that uses comments in conjuction with the ASTs to generate source code summaries using graph neural networks. This is one of the only works we have found that actually addresses the filtering process in detail and specifically addressed (filtered) comments written in a language other than English. However they still assume that the first sentence of a comment suffices for a general summary, in a corpus of 1.9 million method-comment pairs. In addition they only use similarity metrics such as BLEU~\citet{Papineni02bleu:a} scores for assessing the quality of their system (consistent with standard practice for similar work~\cite{LeClairNeuralModel,movshovitz2013natural,awad2019commit,cortes2014automatically,jiang2017automatically,iyer2016summarizing,chen2021my}). 

They never look at other metrics such as as relevance of the generated summary to the source code, helpfulness of the summary to the developers, or actual readability  of the final summary. We show later that even if the BLEU score is high for a particular sentence, it does not necessarily mean that the sentence is actually coherent or useful, consistent with other work showing that high BLEU scores by themselves can be misleading in general~\cite{callison2006re,callison2008further,sulem2018bleu} and specifically for software engineering tasks~\cite{stapleton2020human,roy2021reassessing}, in addition to being highly sensitive to minor details of preprocessing~\cite{post2018call}.

There is interesting prior work on comment quality with respect to utility \emph{for human developers} (the primary reason comments exist at all).  This is an important focus, but this notion of comment quality does not necessarily coincide with usefulness for comments for natural language-based techniques~\cite{pandita:specGen,mcmillan:dataset,neuralModelGen}. We show that some comments (such as those in Figure \ref{fig:mixedcomments}) are useful for developers yet are not suited for building linguistic machine learning models, and therefore need to be pruned during input preparation. 

\citet{javaDocQuality} provide heuristics for the analysis of the comment's quality, however the paper focuses on the qualitative analyses instead. The authors point out that comments are crucial for timely and well-performed software maintenance, however comments are often overlooked by developers and do not contain enough information for various development purposes. They proposed several metrics to analyze the quality of a comment with regard to human use: the number of tokens, nouns and verbs, number of words per Javadoc comment, number of abbreviations, readability and code/comment consistency.

They analyze how well the source is documented and explore the correlation of the code documentation and bugs in the software.  The analysis, however, is limited in the size and scope. The authors only analyzed a small number of related projects and looked only at how well the comments were written. The authors did not provide any insight into the actual semantic contents of the comments. In addition since they were focused on Javadoc comments, they did not evaluate different categories of comments, such as copyrights and other types of comments that exist, but do not help with the documentation of the code for human readers.


\citet{qualityOfComments} analyzed comment quality in Java and C++, bucketing comments into categories such as copyright, header comments, inline comments, method comments, and others. They used machine learning to classify comments into each category with high precision and recall and provided a good metrics for overall human-relevant comment quality, such as usefulness, relevance, completeness and consistency. However, the authors did not explore in detail the contents of all comments, or comment properties that might affect other uses. During their preprocessing they removed special characters and other types of symbols from the comments. As we show later, this is inadequate for extracting high-quality linguistic models from comments, as it leaves significant non-linguistic data.

\begin{figure}[t!]
\begin{subfigure}[b]{0.45\textwidth}
	\centering
\begin{lstlisting}[basicstyle=\scriptsize\ttfamily]
decoding_table = (
	u'\x00'     #  0x00 -> NULL
	u'\x01'     #  0x01 -> START OF HEADING
	u'\x02'     #  0x02 -> START OF TEXT
	u'\x03'     #  0x03 -> END OF TEXT
	u'\x9c'     #  0x04 -> CONTROL
	u'\t'       #  0x05 -> HORIZONTAL TABULATION
	u'\x86'     #  0x06 -> CONTROL
	u'\x7f'     #  0x07 -> DELETE
	u'\x97'     #  0x08 -> CONTROL
	u'\x8d'     #  0x09 -> CONTROL
	u'\x8e'     #  0x0A -> CONTROL
\end{lstlisting}
	\caption{Comments useful to humans but useless for natural language processing in Cpython\protect\footnotemark repo}
	\label{fig:shortComment}
\end{subfigure}
\begin{subfigure}[b]{0.45\textwidth}
	\centering
\begin{lstlisting}[basicstyle=\scriptsize\ttfamily]
r"""
Display the homogeneous components of the mixed form.

The output is either text-formatted (console mode) or
LaTeX-formatted (notebook mode).

EXAMPLES::

    sage: M = Manifold(2, 'M')
    sage: f = M.scalar_field(name='f')
    sage: omega = M.diff_form(1, name='omega')
    sage: eta = M.diff_form(2, name='eta')
    sage: F = M.mixed_form([f, omega, eta], name='F'); F
    Mixed differential form F on the 2-dimensional
     differentiable manifold M
    sage: F.display() # display names of homogeneous components
    F = f + omega + eta

"""
\end{lstlisting}
	\caption{SageMath\protect\footnotemark comment for \lstinline|MixedForm.display|, containing markup and code}
	\label{fig:sageMath}
\end{subfigure}
\caption{Comments containing non-linguistic text}
\label{fig:mixedcomments}
\end{figure}
\addtocounter{footnote}{-1}
\footnotetext{https://github.com/python/cpython/}
\stepcounter{footnote}\footnotetext{https://github.com/sagemath/sage}

\citet{commentClassification} provide additional insight into different types of comments and their classification. Their work is one of the only papers that acknowledge that not all comments serve the same purpose and there are different types of comments encountered in the real world code bases. For example, there are documentation-purpose comments targeted at programmers, such as Javadoc, and there are inline comments which target internal developers --- aimed at different audiences for different purposes. Some comments are reminders, some provide rationale for implementation choices, and some are merely separators of logical blocks. This distinction is important because comments used for different purposes will have different structure and will have different levels of usefulness for machine learning.
The authors categorize comments by their function and provide statistical analysis and breakdown of the categories. However, this work considers only the frequency of each comment category, not consequences of the frequencies.

This work on comment general quality shares two other important limitations: all of it focuses on Java (with one exception also treating C++ comments), and the size of the corpora studies are both small and manually curated to be of high quality.

\citet{pandita:specGen} evaluate on 2500 \emph{sentences}. \citet{qualityOfComments} evaluate on 830 comments drawn from 12 Java projects, and 500 comments drawn from C++ projects.\footnote{Further details on the C++ data are in a MS thesis that is no longer available online.}  \citet{javaDocQuality} evaluate on comments drawn from 3 releases each of 2 large Java projects known to be of reasonably high quality (ArgoUML and Eclipse), for a total of just over 121K comments --- but this is the sum of all comments in each release, including duplicates.

Focusing on high-quality software as exemplars of good commenting practice is useful for identifying major kinds of comments, but side-steps the general issue of also dealing with comments that may be inappropriate for major tasks. These projects also focus exclusively on method-level comments, discarding information that may reside in comments inside methods.


Work on automatically generating one-sentence summaries of Java methods~\cite{neuralModelGen,mcmillan:dataset,leclair2020improved} uses a larger dataset of 2.1 million pairs of Java method bodies with one-sentence comments, produced filtering comment-method pairs from \citet{linstead2009sourcerer} to select only those pairs whose methods are less than 100 word-tokens long, and whose comments are specifically Javadoc comments between 3 and 13 words long. We discuss the trade-offs of this approach in comparison to ours in Section \ref{sec:limitations}.
This resembles a scaling-up of \citet{howard2013automatically}'s approach, extracting the first sentence of every method-level Javadoc comment, yielding 20,199 such sentences, of which they examined a random sample of 150 in depth.
\looseness=-1

By contrast, we evaluate on two Python comment corpora containing just over 3.7 million comments, the first of which is a new large sample of publicly-available GitHub repositories of Python code.
Our data and code are available online\footnote{\url{https://figshare.com/s/e3cd836401ecadae5b88}} and will be archived with a DOI upon acceptance.

\section{Goals and Approach}
We seek to understand the distribution of code comments suitable for \emph{linguistic} processing, as well as the frequency of \emph{non-linguistic} data, and how removing non-linguistic data affects the performance of language models trained on large comment corpora.
The tasks we are interested in include suggesting comment completions, generating method summaries, or generating code from comments or text.  Training machine learning systems for any of these purposes relies on the assumption that code is not only amenable to statistical and neural techniques drawn from natural language processing, but is specifically (primarily) actually human language. 

To this end, we have built two linguistic models for comment completion --- one statistical, one neural --- and trained each four times: for baseline filtering common across any recent work handling code comments on both an existing corpus and a new large corpus, and again on the same corpora after designing and applying mechanisms to specifically filter identified categories of non-linguistic data.  
We focus on comment completion because it isolates the effects of filtering comments (e.g., compared to interactions filtering pairs of code and comments which may have non-trivial correlations).
We evaluate two kinds of linguistic models on two preparations of two corpora, to evaluate the effects of comment filtering across different linguistic modeling techniques, and using both large scale and carefully-curated corpora.
This section details our data collection, characteristics of the new corpus, the details of our analysis process in terms of the research questions answered (in dependency order) for our ultimate investigation. The next section reports on our results.

%
%

\subsection{Data Collection}
\label{sec:dataCollection}
For this project we used comments from publicly available open source projects from GitHub using GitHub RESTful search API. We processed the 840 most popular Python projects (according to the GitHub search API as of June 2021).
In addition, we couple studies on this new corpus with an additional Python corpus, the SRILab Python 150K dataset~\cite{raychev2016probabilistic}\footnote{https://www.sri.inf.ethz.ch/py150}. While our new corpus selected as many of the most popular and currently maintained projects as Github's API would permit us to retrieve, the Python 150K dataset had more restrictive criteria, restricting to only Python files (not necessarily whole projects) with at most 30,000 AST nodes (in service to AST-based program synthesis goals) with permissive licenses. They have files from 8,422 projects in their dataset. Despite that our GitHub dataset contains smaller number of projects, Figure \ref{tab:stats} shows that our dataset actually contains more lines of code and more comments. This disparity is due to the process of acquiring the datasets. While we downloaded top 840 popular projects from GitHub, SRILab had a different selection criteria and represent a size-limit corpus, in addition to a selection by the license type. The official GitHub API allows downloading only 1000 projects, however there are some duplicate projects in the list which are forks of the same project and therefore the total number of unique projects is less. We did not download the forks of the more popular projects as the original projects had higher rating an were first in the download order, they were the only ones downloaded. The larger number of projects in the SRILab dataset can only be acquired by crawling the GitHub web site.

The two datasets overlap on 35 projects, such as several implementations of Python, PyMySql, nltk and others. However our GitHub dataset is several years newer than the SRILab dataset, and contains much more code overall (39\% more files, 90\% more lines of code, and 46\% more unique comments per Tables \ref{tab:comStats} and \ref{tab:stats}), so while the corpora are not completely independent, the overlap is very small.


For the GitHub dataset we were able to open 208,661 files, discarding 90 in an encoding that was not ASCII or any flavor of Unicode.
From those files we extracted 2,436,575 comments prior to filtering, 592,063 of which were Python docstrings.
The SRI Lab dataset contained 149,970 files, from which we extracted 1,276,558 comments, of which 365,083 are docstrings, and were unable to open 28 files which were written in a uncommon encoding.

We implemented custom comment extraction logic, which parses out both Python ``docstrings'' (multi-line string literals which are deemed comments based on their syntactic location) and single-line comments (which standard Python parsers discard during parsing). 

\begin{figure}
	\centering
	\includegraphics[scale=0.71]{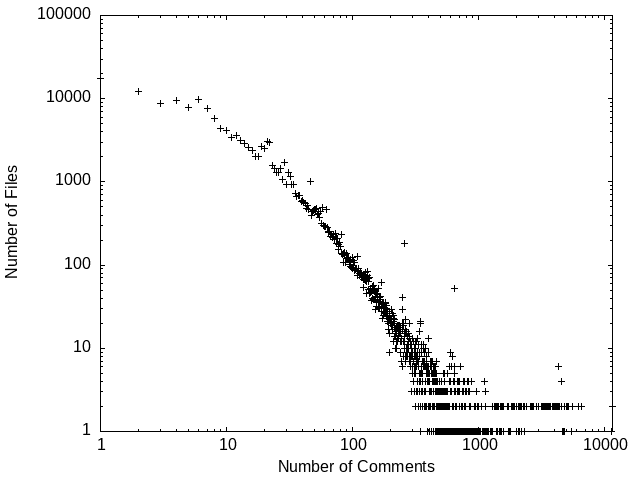}
	\caption{Statistics of how many files contain specific number of comments in Github dataset}
	\label{fig:comPerFileGit}
\end{figure}

\begin{figure}
	\centering
	\includegraphics[scale=0.71]{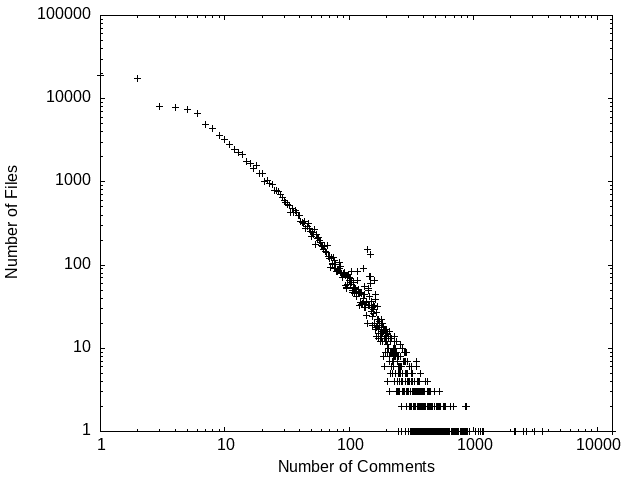}
	\caption{Statistics of how many files contain specific number of comments in SRILab dataset}
	\label{fig:comPerFileSrilab}
\end{figure}

\begin{table}
	\centering
	\caption{Breakdown of the number comments per files and lines of code}
	\begin{tabular}{|r|r|r|r|r|r|r|}
		\hline
		& \multicolumn{3}{c|}{GitHub} & \multicolumn{3}{c|}{SRILab}  \\ \hline
		\# Comts & \# Files & \# LoC & LoC Avg  & \# Files & \# LoC & LoC Avg \\ \hline
		0     &  43909 &  2081111 & 47.40 &  22269 &   979823 & 44.00 \\ \hline
		1     &  17551 &  1646820 & 93.83 &  18903 &  1646895 & 87.12 \\ \hline
		2     &  12127 &  1045279 & 86.19 &  17369 &   891869 & 51.35 \\ \hline
		3     &   8863 &   904753 & 102.08 &   8046 &   752938 & 93.58 \\ \hline
		4     &   9391 &  1041079 & 110.86 &   7794 &   690436 & 88.59 \\ \hline
		5     &   7845 &   875778 & 111.64 &   7384 &   909928 & 123.23 \\ \hline
		6     &   9718 &  1141387 & 117.45 &   6569 &   690690 & 105.14 \\ \hline
		7     &   7726 &  1147669 & 148.55 &   4956 &   576455 & 116.31 \\ \hline
		8     &   5775 &   720700 & 124.80 &   4401 &   523181 & 118.88 \\ \hline
		9     &   4424 &   617946 & 139.68 &   3586 &   472613 & 131.79 \\ \hline
		10    &   4100 &   621704 & 151.64 &   3243 &   435779 & 134.38 \\ \hline
		11-100 &  68890 & 22559732 & 508.44 &  40648 & 10980807 & 439.63 \\ \hline
		100-2000 &   8205 & 12306553 & 3037.54 &   4795 &  4989213 & 2079.67 \\ \hline
		Total & 208524 & 46710511 & 4780.09 & 149963 & 24540627 & 3613.67 \\ \hline
	\end{tabular}
	\label{tab:comStats}
\end{table}

\subsection{Dataset Characteristics}
We give some statistics on our datasets, both to provide a general high-level characterization, and because the statistics suggest the collected body of comments is plausibly representative of general Python code.
\Cref{fig:comPerFileGit,fig:comPerFileSrilab} (note the logarithmic scales on both axes) show that the vast majority of files contain only a small number of comments (less than 10) while a small number of files contain a large number of comments in them, which can also be seen from Table \ref{tab:comStats}. From the figures we can observe that that in addition to containing small number of comments, a majority of the source files are also small on average, containing fewer than 150 lines of code and that in general larger files tend to have more source comments in them.  \figref{comPerFileGit} in addition shows an interesting trend at the bottom of the graph that is not present in \figref{comPerFileSrilab}: a wider distribution of files that contain very large numbers of comments, approaching 10,000. This may be due in part to the SRILab dataset's focus on programs of limited size.

Table \ref{tab:comStats} shows that we separate the files into three groups by the number of comments they contain. We inspected some of the files that belong to each category, which provided us with some insights into what types of files belong to each group. While detailed analysis of each category is subject to further research, we can infer some characteristics of the them.

We observed that the largest group with 10 comments or fewer includes a number of data initialization scripts, configurations, low-level details (e.g. a set of database queries for a higher-level API), and similar code. We found that these files are often very small, containing mostly the object initialization and setup in main functions of programs or scripts. The small amount of comments in these cases is likely due to the generally limited overall file sizes.

The files we observed containing 11-100 comments are generally class definitions for objects that appear to work at a high level of abstraction (i.e., collecting functionality of other units, rather than directly implementing tricky functionality). These examined files commonly contained many comments that at least superficially appear to contain high-quality predominantly-linguistic data. The files containing over 100 comments is the smallest group. The files we inspected in this category largely consisted of library files that are defining large APIs.

The files with many comments, however, do not always contain useful comments. There are several notable outliers in the dataset that contain more than a 1000 comments and more than 5000 lines of code in a single file, with the most notable file containing more than 80,000 lines of code. We did not include these especially large files in the table \ref{tab:comStats} as they are outliers and are not representative of the overall trend in our data. We did, however use these outliers in our dataset. Some of these files contain large map definitions for encoding schemes such as font or character sets, as in \figref{shortComment}. They often contain large numbers of comments, but these comments are not useful for linguistic purposes. Some other files contain definitions of unit tests with thousands of tests in them. These files contain a high degree of similar code and comments. \figref{manticore} shows one such example where the file contains over 1000 comments for unit tests that differ in only 1 or 2 words. The last category of special-case files that we observed in our GitHub dataset are the source code that contains encoded binary objects. These files can have thousands of lines of code, but most of it would be actually encoded binary. These files contain few, if any, comments and are also largely useless for the natural language processing and language-oriented machine learning algorithms.

\begin{figure}[t]
	\centering
	\begin{lstlisting}[basicstyle=\scriptsize\ttfamily]
		def test_AAD_1(self):
		""" Instruction AAD_1
		Groups: not64bitmode
		0x80702ff:	aad	0xff
		"""
	\end{lstlisting}
	\caption{A comment from the Manticore\protect\footnotemark{} project that can't be used for NLP}
	\label{fig:manticore}
\end{figure}
\footnotetext{https://github.com/trailofbits/manticore}

This data shows our corpus has coverage of large and small projects and files, heavily commented and lightly commented, which likely covers a variety of commenting practices.

\subsection{Research Questions}
\label{sec:rqs}
Our ultimate goal is to evaluate the impact of non-linguistic data in comments on machine-learning algorithms processing comments. However, while such an investigation clearly requires removing such non-linguistic data, we initially had no knowledge of how to do this, or even what kinds of non-linguistic data might exist, or what kinds might be the most problematic (i.e., recurring, rather than having a single exceptional example).
Thus we set a series of research questions to answer, in order, to work towards such an evaluation. Each of these research questions depends on the results of the prior research questions.

\begin{rquestion}
	\resq{1}
\end{rquestion}

Building a linguistic model that generates coherent English comments requires training data that is (mostly) valid English prose relevant to source code.
Common sense and prior experience indicate that there are many ways to deviate from that scope.
Machine learning algorithms applied to text will readily process text filled with things a human would immediately recognize as gibberish or at least not natural language, but will often still produce some reasonable-looking outputs given enough valid information. However, a linguistic model trained on poor quality data will generally produce poorer results than one trained on higher-quality data.
Prior work applying machine learning to comments typically either uses a very small hand-groomed data set of high quality comments~\cite{javaDocQuality,commentClassification,qualityOfComments}, or assume that some level of modest filtering is adequate~\cite{mcmillan:dataset}, based on achieving what are considered good results for the intended task~\cite{neuralModelGen}. However, the former assumption is never fully explained (the criteria for the manual filtering are typically stated without detail\footnote{And in some cases~\cite{LeClair:CodeGNN} there are clearly multiple natural languages in the remaining comments.}) and the latter assumption is  typically not validated on the actual data.


We repeatedly manually reviewed random selections of raw comments extracted via the tool of Section \ref{sec:dataCollection} to identify common themes in comments that were clearly not linguistic in nature. We did fix in advance that we wished to focus on English language comments --- widely known to be the most prevalent natural language in open source software development --- so elected to treat non-English natural language as non-linguistic for the purpose of building \emph{English-language} models.

\begin{rquestion}
	\resq{2}
\end{rquestion}
Identifying meaningfully different categories of comments is important, but does not necessarily imply the good and bad comments for generating linguistic models can easily be separated, either for our purposes or in future work seeking to account for non-linguistic data embedded in comments.  Prior work~\cite{commentClassification} has attempted to categorize comments by purpose using a neural network for a fixed set of comment categories.  However, such an approach is difficult to extend: adding a new class of comment requires retraining the entire network based on significant additional training data.  We seek human-readable textual filters to classify comments, which may be individually understood or extended.

For each problematic category of comment, we iteratively developed one or more deterministic, human-auditable filtering criteria to match human-recognizable patterns in those categories. To ensure that each filter only matches comments in the intended category, we ran it on both files where the relevant category had been observed, as well as randomly selected sets of other files, and then manually inspected the data removed and left by the filter. For each category, we added, generalized, or specialized filters for portions of that category until we were unable to (manually) locate additional examples of that category in filtered data or locate misfiltering examples in the data removed.
\looseness=-1

\begin{rquestion}
	\resq{3}
\end{rquestion}
Given an effective filtering method, we can quantify how prevalent certain types of comments are in raw source code. Beyond being of intrinsic interest, this helps evaluate how critical filtering for specific comment classes is (rare categories are unlikely to pollute models significantly), and can contribute to evaluations of confidence in work that does not account for specific kinds of non-linguistic comments. We quantify how many comments match each criteria we filtered for, further categorized by whether a comment was a Docstring or not.

\begin{rquestion}
	\resq{4}
\end{rquestion}
Literature on comment quality tends to assume comments are binary: either useful or useless.  These positive or negative traits are often assumed to be correlated with factors like size of the comment, but there is limited empirical evidence of this.  Moreover, there has been little exploration of why comments might be inappropriate for linguistic models yet still desirable to have in source files.  For the classes of comments we filter due to being non-linguistic (or not purely linguistic), we describe examples of meaningful comments in that category, justifying their existence.

\begin{rquestion}
	\resq{5}
\end{rquestion}
It is well-known that the quality of data used for machine learning algorithms can have major impacts on the quality of the learned model, and it is known that this has had impact on some software engineering work~\cite{dupComments,mcmillan:dataset}.
We evaluate the impact of filtering on two linguistic models for suggesting comment completions:  a 4-gram model and a neural network for sequence prediction.
We evaluated each model on two variants of each corpus:
\begin{enumerate}
	\item after applying \emph{basic filtering} --- roughly, common baseline best practices, removing special characters and punctuation, described in more detail in Section \ref{sec:analysis}
	\item after applying \emph{advanced filtering} --- all filters identified for Research Question 2. Here we observed that the order of filters matter, as for example, the source code filters need to be applied prior to removal of punctuation. We discuss is in more detail in the later section.
\end{enumerate}

As noted earlier, focusing on the task of comment completion allows us to attribute all changes in model performance on filtered data to the content of comments. Alternative evaluation tasks, such as oracle or formal specification generation from comments~\cite{comToSpec,genExceptions,Zhai:Com2Spec} or automatic method summarization~\cite{LeClairNeuralModel,leclair2020improved,iyer2016summarizing,chen2021my,movshovitz2013natural} train on pairs of code and comments, leaving open the possibility that non-linguistic comments may have a statistical relationship with qualities of code they are associated with, which may confound the impact of comment-focused filtering.

\section{Analysis}
\label{sec:analysis}
\subsection{Research Question 1: Comment Categories}
\label{sec:comCats}
The inspection of the contents of the comments yielded a range of non-linguistic comments, including comments containing source code (not always Python, including \LaTeX, HTML, and parser-generator syntax), comments written in a language other than English, containing no dictionary words, or serving other special functions. The special function comments include copyrights, licenses, encoding declarations\footnote{Hints to editors indicating the correct encoding for the file, common in UTF-8 source files.} in the file headers and other service-type comments that serve purposes other than documenting the source.
Our categories are listed together with their filters (from Research Question 2) in Table \ref{tab:filters}.
Some comments were highly structured, such as the example in \figref{sageMath}, which includes text along with a structured portion specifying intended use of the API.
We also found that many comments contain markup for \LaTeX or Markdown that specify mathematical formulas and in some cases text layout, in addition to the expected non-English human languages. While \LaTeX can be identified by it's keywords and markup commands, the Markdown is often difficult to distinguish in automated way from ordinary comments due to how it is structured. One important property that Markdown contributes to comments, however, is the fact that it produces more short-length comments that do not necessarily contribute much to the linguistic richness of a comment.

\paragraph{Code/Math}
The most commonly observed class of non-linguistic comments contained source code and/or mathematical equations, sometimes mixed with regular linguistic comments.
In cases where code or formulas occur embedded within linguistic text, the surrounding text likely does have some value to linguistic models. However, this text often contains mentions of subformulas or variables, so feeding those into training for linguistic models is of unclear impact.
Comments containing code were sometimes clearly intended to document behavior, but in many other cases appeared to be simply commented-out code left in the repository.
Mathematical equations likely contribute useful information for human readers, but are of no value to linguistic models of comment text. 

We decided to group these two large categories together due to their similarity in representing some technical notation and because many comments contain both, as in Figure \ref{fig:sageMath}. In some cases a comment might contain \LaTeX{} which formats a math equation together with the source code example for this equation.
We have subcategorized this group according to the type of code or math they contained, highlighted in Table \ref{tab:filters}.
\looseness=-1

\paragraph{Non-English} Many comments consist primarily of non-English words/characters.  Comments in other languages have clear value to linguistic models for those languages, but our work is limited to English.
We observed comments in French, Italian, Chinese, Spanish, and other languages.

\paragraph{Copyright} A significant number of comments simply state copyright or license information for a file. As these comments do not actually describe the source code, we are not interesting in them. Moreover, they are often identical across most files in a codebase. Feeding this sort of repetition to a linguistic model would obviously skew the model towards copyright and license statements.

\paragraph{Non-linguistic} Some of the comments consist only of punctuation symbols (e.g. separator comments) or contain ASCII art. Some of these symbols are not part of the standard punctuation removal (e.g. \u{\P} \textltshade \textlfblock \textSFxxiv all appeared in the comments). 

\paragraph{Other} There are also many file encoding directives extracted as comments, for example several forms of \texttt{\# -*- coding: utf-8 -*-} are common. These type of comments provide no linguistic value and 
Other examples include commented out ASCII hash values, or encodings of text in a non-readable format.

\paragraph{Duplicates}
Even from manual inspection, setting aside copyright and licensing statements, it was readily apparent that there were many duplicate comments. 

\subsection{Research Question 2: Effective Filtering}

\begin{sidewaystable}
	\vspace{0.7\textheight}
	\centering
	\caption{Filtering Comment Categories}
	\begin{tabular}{|l|l|l|p{0.4\textwidth}|}
		\hline
		Category & Subcategory & Pattern & Python Regular Expression OR Condition\\ \hline
		\hline
		Copyright & - & case-insensitive match for copyright & \lstinline|if "copyright" in comment:|\\ \hline
		\multirow{9}{*}{Code/Math} & Call site & Matches \ldots\emph{obj}.foo(\ldots)\ldots etc. & \lstinline-'[\w\d]+\s*(={1,2}|\.?)\s*[\w\d_]+\.?[\w\d_]+\(.*\)'-\\
		& Assignment & Matches assignments and comparisons & \lstinline-'[\w\d]+\s*={1,2}\s*[\d]+'- \\
		& Hash Values & Match 32- or 64-char hex strings & \lstinline|'[a-f0-9]{32,64}'|\\
		& LaTeX & Common LaTeX math sequences &
		\begin{minipage}{0.5\textwidth}
			\begin{lstlisting}
'(\\begin\{\w+\})|(\{?\\(alpha|beta|gamma|omega|lambda)\}?)|
				(\\\\?mathbf\{[\w]{0,5}\})'
			\end{lstlisting}
		\end{minipage}
		\\
		& SageMath & SageMath interactive excerpts & \lstinline-'sage:\s*($\mathit{callsite\_pattern}$)?'-\\
		& HTML & Strip HTML/XML tags & \lstinline-'<[^>]*>'-\\
		& Antlr & Antlr parser-generator docs & \lstinline-'\$ANTLR|type:'-\\ \hline
		Non-English & - & Dictionary \& classifier & See prose description below\\ \hline
	\end{tabular}
	\label{tab:filters}
	\bigskip\bigskip
	\caption{Comment Categories}
	\label{tab:stats}
	\begin{tabular}{|l|rrrr|llll|}
		\hline
		& \multicolumn{4}{l|}{GitHub}                                                                                & \multicolumn{4}{l|}{SRILab}                                                                               \\ \hline
		& \multicolumn{2}{c|}{With duplicates}                           & \multicolumn{2}{c|}{Without Duplicates}   & \multicolumn{2}{l|}{With Duplicates}                           & \multicolumn{2}{l|}{Without Duplicates}  \\ \hline
		& \multicolumn{1}{r|}{All}     & \multicolumn{1}{r|}{Docstrings} & \multicolumn{1}{r|}{All}     & Docstrings & \multicolumn{1}{l|}{All}     & \multicolumn{1}{l|}{Docstrings} & \multicolumn{1}{l|}{All}    & Docstrings \\ \hline
		Code/Math      & \multicolumn{1}{r|}{320258}  & \multicolumn{1}{r|}{186723}     & \multicolumn{1}{r|}{229544}  & 126258     & \multicolumn{1}{l|}{176770}  & \multicolumn{1}{l|}{170128}     & \multicolumn{1}{l|}{133732} & 132693     \\ \hline
		Non-English    & \multicolumn{1}{r|}{74020}   & \multicolumn{1}{r|}{3088}       & \multicolumn{1}{r|}{22332}   & 7408       & \multicolumn{1}{l|}{2251}    & \multicolumn{1}{l|}{12}         & \multicolumn{1}{l|}{1623}   & 12         \\ \hline
		Copyright      & \multicolumn{1}{r|}{66118}   & \multicolumn{1}{r|}{4494}       & \multicolumn{1}{r|}{7627}    & 1911       & \multicolumn{1}{l|}{35179}   & \multicolumn{1}{l|}{6432}       & \multicolumn{1}{l|}{9661}   & 3168       \\ \hline
		Non-Linguistic & \multicolumn{1}{r|}{60855}   & \multicolumn{1}{r|}{5789}       & \multicolumn{1}{r|}{9509}    & 2923       & \multicolumn{1}{l|}{2098}    & \multicolumn{1}{l|}{1067}       & \multicolumn{1}{l|}{1030}   & 864        \\ \hline
		Duplicates     & \multicolumn{1}{r|}{1216476} & \multicolumn{1}{r|}{682986}     & \multicolumn{1}{r|}{0}       & 0          & \multicolumn{1}{l|}{445693}  & \multicolumn{1}{l|}{228885}     & \multicolumn{1}{l|}{0}      & 0          \\ \hline
		Latex          & \multicolumn{1}{r|}{1929}    & \multicolumn{1}{r|}{1313}       & \multicolumn{1}{r|}{1663}    & 1047       & \multicolumn{1}{l|}{210}     & \multicolumn{1}{l|}{388}        & \multicolumn{1}{l|}{204}    & 364        \\ \hline
		HTML           & \multicolumn{1}{r|}{76594}   & \multicolumn{1}{r|}{40249}      & \multicolumn{1}{r|}{46965}   & 25042      & \multicolumn{1}{l|}{21762}   & \multicolumn{1}{l|}{21862}      & \multicolumn{1}{l|}{14792}  & 15861      \\ \hline
		Total          & \multicolumn{1}{r|}{2473403} & \multicolumn{1}{r|}{1493669}    & \multicolumn{1}{r|}{1256927} & 810684     & \multicolumn{1}{l|}{1302462} & \multicolumn{1}{l|}{781975}     & \multicolumn{1}{l|}{856769} & 550269     \\ \hline
	\end{tabular}	
\end{sidewaystable}

For each filterable category, the set of filters we developed are shown in Table \ref{tab:filters}.
To identify (typically, Python) code within comments our filters look for patterns such as assignments and method calls.
We also used separate patterns to match a variety of formal languages nested inside comments: HTML, ANTLR~\cite{parr1995antlr} parser-generator comments (indicating generated code), \LaTeX, and prefixes for the way SageMath represents example interactive Python sessions in comments. We also match 32- or 64-character hexadecimal strings, which typically indicate hashes of objects outside the program that are not useful for linguistic models, like git commit hashes, IPv6 addresses, or hard-coded smart contract addresses.

To identify a non-English text we used a combination of checking for non-ASCII characters (though this is not tenable for languages whose writing systems use accents, umlauts, diareses, etc.), building a classifier with the NLTK~\cite{nltk} toolkit for implementing NLP tasks in Python, and using the \texttt{langid.py}~\cite{langid} natural language classifier. Where \texttt{langid.py} was unable to give clear classification, we complemented its results by checking with the English language dictionary from NLTK, and checks on character encodings.  This filter kept only comments where all 4-grams (subsequences of 4 words) were classified as English (\texttt{lang.py} is significantly more accurate on fragments than individual words).
\looseness=-1

A separate discussion is warranted in regards to the order of the filter application. As filtering is destructive by nature, applying a complex set of filters also requires special ordering in case they overlap. For example, removing punctuation (a standard pre-processing step for training models on text) will remove some of the indication that a comment contains code or technical markup. As we can see from table \ref{tab:stats}, source code is one of the largest prevalent category of non-linguistic type of comment, and thus failing to remove it will have a large impact on the resulting data. Another example where the order of the filters matter is the encoding directive  \texttt{\# -*- coding: utf-8 -*-}. Here, stripping the punctuation first reduces the size of the comment to two words, which will be removed by the length filter in the next step. Reversing the filtering order in this case will result in failing to filter this directive.

There was one category of non-linguistic comments that we could not effectively filter: license clauses.
In contrast to copyright statements, which surprisingly can be filtered with high accuracy by simply looking for the word "copyright," we found no effective means of filtering all (or even most) and only license clauses.  All license clauses we identified include the word ``license'' but this alone is too coarse a filtering criteria, removing many legitimate comments (e.g., dealing with software usage licensing).

\subsection{Research Question 3: Category Prevalence}


Based on the filters developed for the previous research question, Table \ref{tab:stats} gives counts of how many comments fall into each category. The numbers include duplicates (i.e., the number of Docstrings containing copyright statements includes duplicates). We have also encountered a total of 118 files that could not be easily opened because the files were not in a typical \texttt{latin1} or unicode encoding. After inspecting these files in our dataset, they did not contain any useful comments, so we chose to ignore them. However, this is also an important factor because some comments within a code base might not be encoded in the same encoding as the whole file and will produce gibberish text if processed as-is: common recommendations for text processing in Python suggest opening the file with a specified encoding and a directive to ignore bytes that do not conform to that encoding,\footnote{e.g., \url{https://stackoverflow.com/a/46781624}} which can result in missing characters that could be important for accurate processing of comments (or in other work, code).

\subsection{Research Question 4: Value of Non-Linguistic Comments}
\label{sec:rq4}
For some categories identified in previous research questions, the need is obvious: license headers dictate how code may be reused, copyright headers indicate the claimed copyright owner. Non-English comments presumably have similar uses to English linguistic comments, but for those reading the appropriate language. Strictly speaking non-English comments deserve further follow-up, as given the English-centric nature of computing, it is reasonable to expect non-English comments to at least seek to meet some additional needs like translation or re-explanation~\cite{guo2018non,dasgupta2017learning}.

This leaves the code and math category. Discussion in Section \ref{sec:comCats} explained that comments belonging to these categories frequently overlap. 
A good example of this is shown in Figure \ref{fig:sageMath}, which shows a comment from the SageMath\footnote{\url{https://www.sagemath.org/}} project,  is a large library used for computational mathematics. Specifically, the figure shows the start of the \lstinline|display| method on the \lstinline|MixedForm| class used to represent mixed forms of differentiable manifolds. (The comment text has been slightly re-flowed to fit in the column, in ways that do not affect significant whitespace.)

This comment shows a number of interacting features of specialized comments. First, this comment contains both Python code snippets and a mathematical formula.  Moreover, both code and formula are \emph{nested} within the larger comment whose overall structure is dictated by the Sphinx\footnote{\url{https://www.sphinx-doc.org/}} 
documentation tool, which assumes comments are ReST (ReStructured Text)\footnote{\url{https://docutils.sourceforge.io/rst.html}}
formatted (akin to Markdown) with sections specific to the SageMath project.  The build process for SageMath processes the comment above into visually-appealing API documentation\footnote{\url{https://doc.sagemath.org/html/en/reference/manifolds/sage/manifolds/differentiable/mixed_form.html\#sage.manifolds.differentiable.mixed_form.MixedForm.display}}
including syntax highlighting for the embedded code sample.
Note that Python, unlike Java, has no language-standardized equivalent of Javadoc which all comments may be assumed to be written in; Sphinx is an independent project from Python itself. Other examples are more complex, containing additional documentation sections, and even embedded LaTeX math markup.

\begin{figure}[t]
	\centering
	\includegraphics[scale=0.5]{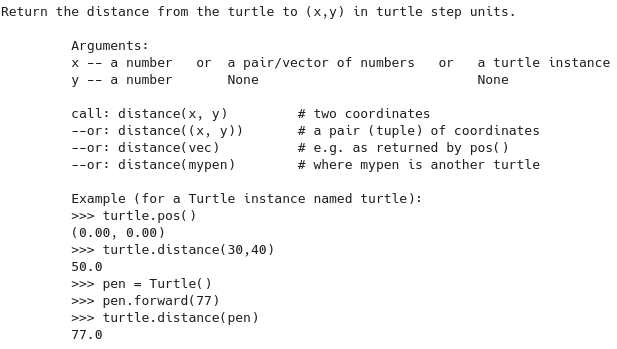}
	\caption{Example of a good comment that can't be used for text generation\protect\footnotemark}
	\label{fig:uselessComment}
\end{figure}

SageMath may be an outlier in this regard: it is one of the largest Python projects in our corpus, and with a userbase consisting largely of professional mathematicians who also write Python code, it may not be representative of Python developer practices in general.
However, even Python's standard library contains examples with some kinds of structure, such as Figure \ref{fig:uselessComment}'s documentation of a method in an implementation of LOGO~\cite{papert1972teaching} in Python.

\footnotetext{https://github.com/python/cpython/blob/master/Lib/turtle.py}

In both of these cases, and many more, the docstrings provide good descriptions of the relevant method, its arguments, and (via examples formatted as embedded code) its usage.  This is essential information for high quality comments~\cite{javaDocQuality,commentClassification,qualityOfComments}.  However, these comments are not directly useful for linguistic models; they contain some short snippets of text, but they are intermingled with more structured non-linguistic markup or ad hoc formatting. Separating mixed natural and formal language is a distinct, heavily studied, open problem in computational linguistics~\cite{wolska-kruijff-korbayova-2004-analysis}.  Even parsing mixed structured input languages that can be combined is a difficult problem~\cite{barrett2016fine}.
This suggests that making effective use of structured comments like these in training linguistic models, or further, training systems that can understand these structured comments without being explicitly coded for the particular structure (e.g., Sphinx comments), represents a significant research challenge. 
There are more of these than duplicates in raw data, suggesting this is a challenge worth tackling.
\looseness=-1

\subsection{Research Question 5: Filtering Impact}

It is well-known that data quality can significantly impact the results of machine learning algorithms, so in practice any attempt to use machine learning will include some baseline cleanup of major issues (i.e., ones that prevent even providing input to the algorithm).  However, given that some prior work applying ML to large code samples did not even address issues with duplication (\citet{dupComments} gives examples and a comprehensive study), and the establishment of a general taxonomy of comment types is an active topic of research~\cite{commentClassification}, it is unclear what baseline filtering of comments might look like.
\looseness=-1

For this reason, we implemented the two levels of filtering mentioned in Section \ref{sec:rqs}, and evaluated two machine learning techniques on two filtered datasets:

\begin{itemize}
	\item \textbf{Basic Filtering} has been filtered with minimal best practices, and removes comments which would cause problems with some tokenization approaches. The basic filtering corresponds to standard data cleaning practices such as normalizing text to lowercase, removing punctuation (including deleting URLs and stripping HTML tags, removing extra blank lines and extra white space, removing some special symbols (also including comments that contain nothing but symbols), and removing duplicates. We also removed comments that were fewer than 10 characters in length or fewer than 4 words, and all comments matching the copyright pattern. Just doing this filtering reduced the number of comments from our Github corpus to 1,652,974, a 66.5\% reduction. However, as our results show, this result is far from being a properly cleaned dataset. Table \ref{tab:comStats} shows that in addition to just punctuation, the comments contain a number of code snippets and non-English comments, which this filtering ignores.
	
	\item \textbf{Advanced Filtering} is first normalized to lowercase, then comments matching any of the patterns in Table \ref{tab:filters} are dropped, then URLs and miscellaneous symbols are removed. Applying filters prior to general removal of punctuation is crucial because many filters make critical use of punctuation.\footnote{Note that this means the first pre-processing step applied by most NLP-with-comments work removes information that is important for filtering non-linguistic data.}
	After this filtering was performed the Github dataset was further reduced by 22.7\% compared to the basic filtering (74.2\% compared to the raw data), with 1,277,176 comments remaining.
\end{itemize}

We do not claim that what is here called ``advanced filtering'' is optimal for removing all non-linguistic comments, even for our dataset (see our discussion of limitations in Section \ref{sec:discussion}).  A deeper study of reliable filtering methods for more complete comment type taxonomies~\cite{commentClassification} could improve the quality of our data even further.  The term \emph{advanced} refers to the level of sophistication compared to basic filtering. 


To evaluate the impact of this filtering, we evaluated the quality of two kinds of comment completion models on each level of filtering (basic, advanced) for each corpus (our Github corpus, and the SRILab Py150 corpus).
First, we built linguistic models with Python Natural Language Toolkit~\cite{nltk} using 4-grams with the Naive Bayesian~\cite{manning99foundations} probability distribution, as n-gram performance is well understood and the model can be easily explored.
Second, we implemented a modestly-sized neural network tailored for sequence prediction.
As mentioned earlier, we chose comment completion as a task because it depends exclusively on comment filtering (and not the quality of some adjoined dataset, like corresponding code fragments~\cite{mcmillan:dataset}), and because it depends largely on the linguistic coherence of the data (as opposed to depending on non-linguistic but otherwise useful comments). For each of the datasets and each filtering approach we extracted a set of 10,000 sentences at random prior to the learning process. We extracted prefixes from these sentences to use for evaluating the comment completion models. The same set of prefixes was used for the corresponding 4-gram model and neural network model.

\subsubsection{4-gram Linguistic Modeling}
In the experiment we generated 10,000 sentences from each of the models. The generation process was the same for each model: for every prefix search for the next word until either no words are left or the produced sentence is longer than 40  words (as otherwise the model would generate infinite sequences for some seeds). The prefixes were randomly selected from each dataset and the sentences they were extracted from were not included in the learning process. Because of this experiment we only wanted to see only the highest quality sentences, we only selected matches such that $p(w_4|w_1,w_2,w_3) >= 0.9$, where $p$ is posterior probability and $w_i$ are words in the order they appear in a sentence. If the model prediction for the next word was less than that threshold, generation was stopped.

For each model, we classified how many of the 10,000 generation attempts produced sentences of length 4 (i.e., how many sentences the model generated no extension for), how many contained code or formula fragments, and how many were grammatically correct according to an English-language parser.
Our results are shown in Table \ref{tab:genStats}.

\begin{table}
	\centering	
	\caption{Statistics for the generated sentences for each model. The \textit{\% 1} column shows the percentage of the Bleu scores of 1 in specific categories}
	\label{tab:genStats}
	\begin{tabular}{|l|l|l|l|l|l|l|l|l|} 
		\hline
		& \multicolumn{4}{l|}{GitHub           } & \multicolumn{4}{l|}{SriLab           }  \\ 
		\hline
		Category        & Basic  & \%~ 1 & Advanced  & \%~ 1     & Basic  & \%~ 1 & Advanced  & \%~ 1      \\ 
		\hline
		Length 4        & 612    & 90.2  & 353       & 81        & 353    & 87    & 436       & 79.8       \\ 
		\hline
		Average Length  & 16     & -     & 17.18     & -         & 15.52  & -     & 16.65     & -          \\ 
		\hline
		Code Artifacts  & 804    & 19.7  & 94        & 23.8      & 678    & 22.5  & 82        & 18         \\ 
		\hline
		Non-English     & 39     & 37.5  & 0         & -         & 6      & 16    & 0         & -          \\ 
		\hline
		Unparsed        & 95     & -     & 13        & -         & 62     & -     & 11        & -          \\
		\hline
	\end{tabular}
\end{table}

The categories in the table are the following:
\begin{itemize}
	\item \textbf{Length 4:} the number of generated sentences of length 4.
	\item \textbf{Code Artifacts:} the number of sentences, containing  elements that came from code or an equation.
	\item \textbf{Non-English}: the number of generated sentences with non-English words in them, or an entirely non-English sentence. 
	\item \textbf{Unparsed:} the number of sentences that the were grammatically incorrect and were not able to be parsed by a parser for English language.
\end{itemize}
If the generated sentence is of length 4, it means that the model could not generate anything for a given prefix. However, it is needed to note that some of the prefixes are grammatically complete phrases and would not necessarily need to have any additional words generated for them. For example both GitHub models produced a phrase ``unpopular query goes viral'' while both SriLab models generated sentences ``parameters for removing interfaces'' and ``call after posting auth.'' All of these are prefixes given to the models and while not being proper English sentences, can both serve as comments for the source code.  On the other hand, there are many cases where the basic model generates a nonsensical sentence and the model with the advanced-filtered dataset generates a somewhat readable sentence. For example (prompt in italics):

\begin{itemize}
	\item \textbf{Github Basic:} \emph{it can be used} with tfdatadatasetmap to apply class weighting
	\item \textbf{Github Advanced:} \emph{it can be used} to read a specific corpus format
\end{itemize}

The Code Artifacts category is of special interest here and it also has some overlap with the sentences of length 4. As the models are trained on the filtered datasets, all punctuation and the structure (formatting of comments that can be code-like with many spaces and tabs in them) are removed. Therefore it is very hard to tell if a particular sentences actually was a part of the source code or an equation or not. However, we can still see artifacts that appear in many of these sentences that can show that they came from a code or an equation. Examples of these artifacts would be strings of the form \texttt{x 5}, which likely came from the assignment \texttt{x = 5}. Another example of an artifact are composite words that could be either variable or function names like \texttt{classplexapiphotophotoalbum}, which came from the comment \texttt{Return the photo's :class:`~plexapi.photo.Photoalbum`}. Strictly speaking, this is not a code, but a reference to a named object, therefore we did not explicitly filter these kinds of comments even during the advanced filtering step. As a result, the linguistic model with advanced filters applied has some artifacts as well, which came from variable names and named objects. These items can be processed further with more advanced NLP methods and add more semantical knowledge to language model. For example, \cite{Hill:2010:Diss} uses named objects and function names in their generation process.

In addition we can see that having even small numbers of non-English words in the dataset with the basic filtering applied will have impact on the training model as we have still generated several sentences with non-English words in them by randomly sampling the dataset.
\looseness=-1

As a coarse automated measure of text quality, we attempted to parse each generated completion with a modern English-language parser. We chose the \texttt{depccg} parser~\cite{yoshikawa:2017acl} that was trained on CCGBank~\cite{hockenmaier2006creating} --- a large natural language corpus from the Wall Street Journal, with words hand-annotated for part of speech~\cite{Marcus:1993}. This parser tends to only parse \emph{grammatically correct sentences}, unlike statistical parsers like Stanford NLP Stanza~\cite{qi2020stanza}. As table \ref{tab:genStats} shows, the models trained after advanced filtering produce more grammatical sentences.
This does not imply the sentences were semantically sensible, but that is much more difficult to evalute.


Some sentences in the basic model were parsed completely even though they contain source code artifacts. The main reason for this is that most modern parsers use probabilistic parsing models and have some room for parsing never-seen before words. In case a sentence contains one or two unknown words to the parser, it can probabilistically assign a role to these words (e.g. a noun or an adjective). Consider:
\begin{itemize}
	\item \textbf{S1:}  \texttt{
		name attrs angle anglea90angleb0rad00 angle3 anglea90angleb0 
		arc \\anglea0angleb0armanonearmbnonerad00 arc3 rad00 bar 
		\\arma00armb00fraction03anglenone 
		note that 3 6 2 8 8 0 0 0 0 0 0 0 0 0 0 0 0 0 0 0 0 0 0 0 0 0 0
	}
	\item \textbf{S2:} \texttt{rabstract base class for 2d lines using the estimated model\\ 
		modelclsresidualsdata all data samples with residuals smaller than the maxlen}
\end{itemize}

S1 is a sentence that was not able to be parsed while S2 is a sentence that was parsed, despite containing several non-words.

\begin{figure}[h]
	\centering
	\includegraphics[scale=0.43]{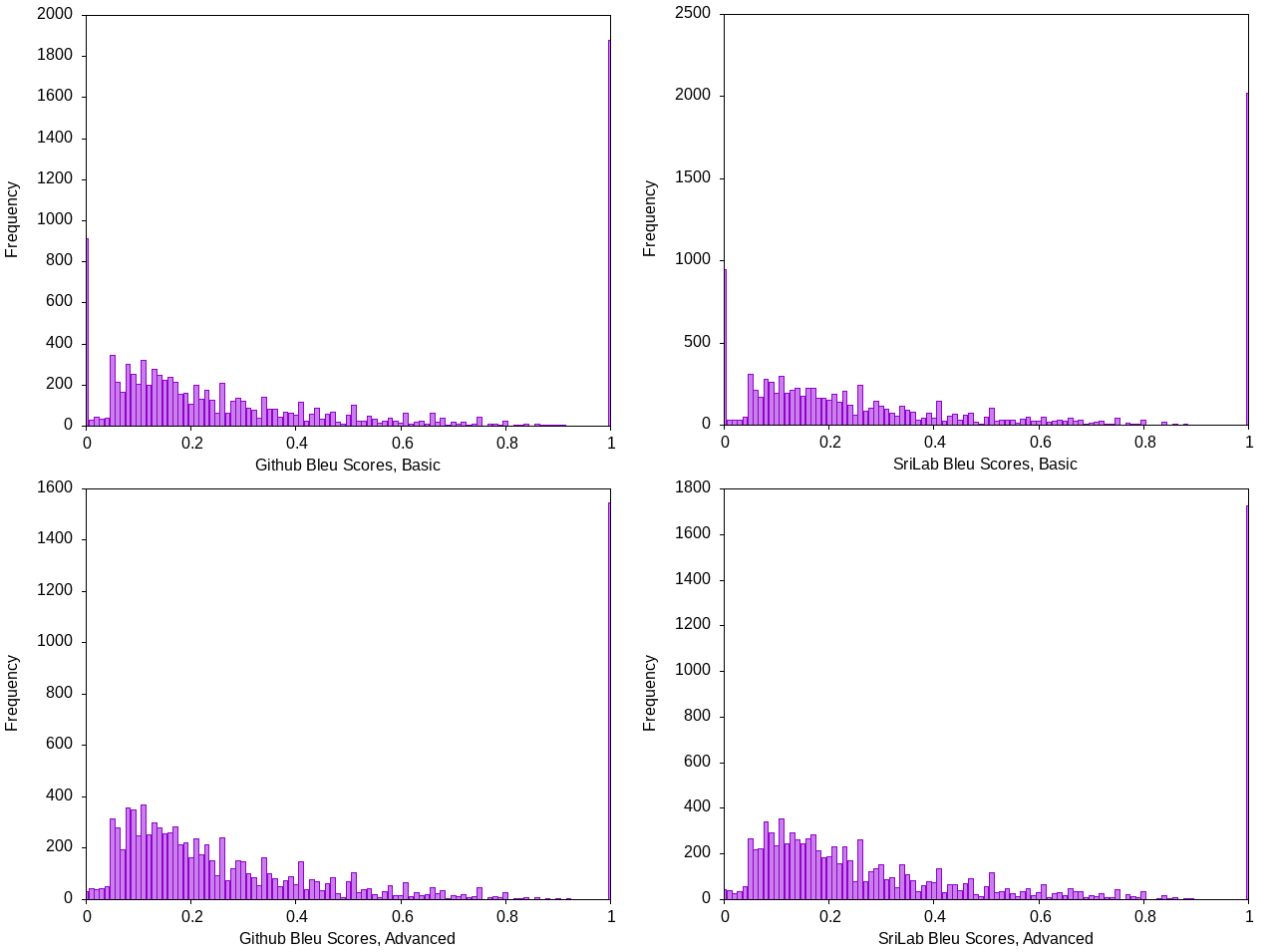}
	\caption{Distribution of the BLEU scores for all datasets.}
	\label{fig:allBleu}
\end{figure}

Finally, as another means to assess overall model performance, we computed 4-BLEU scores~\cite{Papineni02bleu:a}, which we produced using the NLTK toolkit. This is a metric commonly used for evaluating the quality of predicted text compared to a baseline, frequently used for evaluating automatic comment generation~\cite{LeClairNeuralModel,chen2021my}. The scores for all models for GitHub and SRILab datasets can be found in \figref{allBleu}. From the figures we can observe that the models produced from the basic filtering datasets generate many sentences with BLEU score 0, indicating the result was highly textually dissimilar from the original comment. However the models produced from the advanced filtered datasets have few with 0 scores, and the average BLEU score of the generated sentences is higher. Note the y axis of each plot differs. Another observation in the data that the total number of sentences with the BLEU score 1 went down slightly in the advanced filtered dataset. This, however, does not necessarily mean that the overall quality of the comments went down. There are several contributing factors to this phenomenon, some of which can be observed in the Table \ref{tab:genStats}. One of the major contributor to the BLEU scores of 1 are the sentences of length 4 while the other are sentences with code artifacts in them. 
The ``\% 1'' columns in Table \ref{tab:genStats} give the percentage of results in that category (for counting categories) that had a 4-BLEU score of 1, meaning it perfectly predicted the original comment.

We can see that in our datasets short generated comments of length 4 tend to match the original comment. The raw datasets contain 6.7\% and 7.2\% of comments of length 4 for GitHub and SriLab respectively. Additionally, the average length of the comments across all datasets converges to 9 words. We can see from table \ref{tab:genStats} all models generate comments that on aeverage are twice as long as the average comments in the raw dataset. This also explains why shorter comments will have higher BLEU scores and why so many of the generated sentences of length 4 have a BLEU score of 1. As our 4-gram linguistic model tends to generate longer sentences than the raw dataset, the BLEU scores for it will also drop.

Code artifacts also contribute greatly to the set of sentences with BLEU score 1. Even though only about 20\% of them have a BLEU score 1, because of their number, each contributes several hundred sentences to the count. As advanced filtering remove majority of the source code and source code artifacts, the number of sentences with BLEU score 1 will go down as well.

Combining the two facts above we can explain why better filtering can generate lower BLEU scores: it produces longer sentences and generates fewer sentences with unique words like non-English words or code artifacts. However, the generated sentences are also of a higher quality, and upon visual inspection most of them can be read and understood, which is not the case for the model with only basline filtering.

\subsubsection{Deep Learning Evaluation}

We also trained a small neural network to predict the next word based on up to 10 previous words.
For each level of filtering (basic and advanced), we trained a neural network implemented in Keras, encoding the top 10,000 words in each dataset (plus an `unknown' token, and tokens for the start and end of a sentence). Thus unlike the 4-gram model, this model does not perfectly memorize the vocabulary of its training data.
The network consisted of 5 layers:
\begin{itemize}
    \item A 32-dimensional embedding layer
    \item A GRU~\cite{cho-etal-2014-learning}\footnote{Like the slightly better-known LSTM~\cite{hochreiter1997long}, GRU layers are for long range dependencies. GRUs generally train faster with similar performance to LSTMs.} layer returning sequences
    \item A GRU layer not returning sequences
    \item A dropout layer (20\% dropout rate)
    \item A dense layer of size 10,003, using softmax activation
\end{itemize}
This particular network ends up with 662,932 tunable parameters.
The network was trained with the Adamax~\cite{adamax} optimizer, using categorical crossentropy as the loss measure, for a maximum of 150 epochs, but set to stop training if three consecutive epochs failed to yield improvement.

For each dataset (corpus subject to a given filtering level), the model was trained on a randomly sampled 70\% of subsequences ending in a word in the dictionary (top 10,000 tokens), another 15\% was used for validation, and the remaining 15\% was used for testing.


\begin{table}[t]
    \caption{DL Training on Prediction}
    \centering

    \begin{tabular}{|l||c|c|c|c|c|}
        \hline
                            & \multicolumn{2}{c}{SRI Lab} & \multicolumn{2}{|c|}{Github}\\\hline
        Dataset             & Basic & Advanced & Basic     & Advanced \\ \hline\hline
        Productive Epochs           &143     & 105    & 86    & 112\\ \hline
        Final Training Accuracy     &22.54\%&20.53\%&23.81\%&20.86\% \\ \hline
        Final Validation Accuracy   &23.23\%&21.35\%&24.65\%&21.73\% \\ \hline
        Test Accuracy               &23.34\%&21.27\%&24.56\%&21.64\% \\ \hline
    \end{tabular}
    \label{tab:dl}
\end{table}
Table \ref{tab:dl} reports statistics for each dataset: the number of productive training epochs, and accuracy of the model on the final round of training, final round of validation, and on the test set.

\begin{table}
	\centering
	\caption{Statistics for the generated sentences for each model}
	\label{tab:genNNStats}
	\begin{tabular}{|l|l|l|l|l|l|l|l|l|} 
		\hline
		& \multicolumn{4}{l|}{GitHub           } & \multicolumn{4}{l|}{SriLab           }  \\ 
		\hline
		Category        & Basic  & \%~ 1 & Advanced  & \%~ 1     & Basic  & \%~ 1 & Advanced  & \%~ 1      \\ 
		\hline
		Length 4        & 2362   & 22.8  & 2142      & 14.4      & 2394   & 24.1  & 2439      & 15.4       \\ 
		\hline
		Average Length  & 8.7    & -     & 9         & -         & 8.5    & -     & 7.5       & -          \\ 
		\hline
		Code Artifacts           & 179    & 9.4   & 36        & 7.7       & 174    & 16.1  & 26        & 10.3       \\ 
		\hline
		Non-English     & 28     & 0     & 0         & -         & 3      & 0     & 0         & -          \\ 
		\hline
		Unparsed        & 14     & -     & 12        & -         & 9      & -     & 5         & -          \\
		\hline
	\end{tabular}
\end{table}

We repeated the same evaluation on completing 10,000 sentences as for the 4-gram models, using the same 10,000 prefixes for each model as were used for the corresponding 4-gram model trained on the same data.
Table \ref{tab:genNNStats} shows statistics for the neural network model analagous to the 4-gram results in Table \ref{tab:genStats}. We can see a few notable changes by using different generation model. For one, there are significantly more short comments of length 4. In addition, the neural network models on average generates sentences that are closer to the average length of comments in the original dataset. We also see a large reduction in generated code artifacts from the same datasets.

However, we also see that a total number of BLEU scores 1 went down across all categories. There are several reasons for this. The neural-network-generated sentences contain more four-word sentences than the original source comments. For example the basic filtered GitHub sample contains only 696 sentences of length 4. Which means that the neural network failed to generate longer sentences for majority of the results. The situation reverses for the source code artifacts - the original datasets contain more sentences contain more source code artifacts compared to what was generated by the neural network.

There two most likely reasons for this and both are contributing the behavior we observe. One is that neural network does not work on probabilities and selects the next word in a different manner, compared to a 4-gram model. The second reason is that neural network limits it's vocabulary during the learning process. Due to the fact that we do not lemmatize words in the learning process for the neural network model, it treats all forms of the same word as a different word. It is less of a problem in the 4-gramm model because it utilizes the full lexicon that was provided in the source dataset. However, for the neural network model the quality of the original source plays a significant role. We retrained neural network with the lexicon twice as large (20,000 words rather than 10,000) on one of the datasets and the average generated sentence length only increased by one word, the number of four-word sentences increased by two hundred and the number of generated code went down a little. 

The difference in the way the lexicon is stored between two models plays a large role in this case. The 4-gram model is biased towards low-frequency words in two ways. On one hand low-frequency words will have a lower chance to be generated if they appear as a derivation candidate. However, if they appear in the prefix, then they will increase the likelihood that a specific sentence will be generated. This is the reason why the BLEU scores in general are higher for the 4-gram model: the unique code artifacts and other words in the prefix that are not part of a standard English language cause the generated sentences to resemble the original more often. In the neural network case, however, low frequency words are just dropped, causing the model to lose many candidate words. In return, however, we can see that the number of unparsed sentences went down significantly and overall the sentences generated by the neural network model are more grammatically correct than the ones generated by the 4-gram model. Therefore properly filtering the dataset and reducing the number of the rare words will cause a large increase in the overall quality of the generated sentences in this model.

\begin{figure}[t]
    \centering
    \includegraphics[scale=0.5]{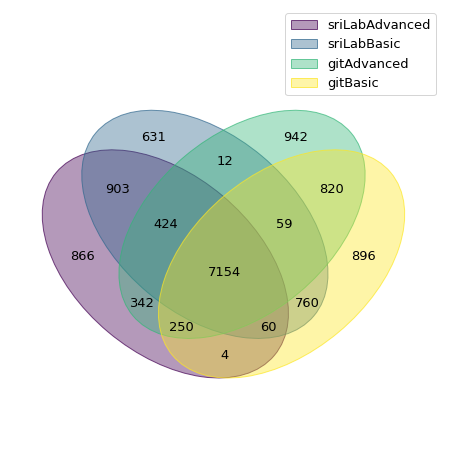}
    \caption{Venn diagram of dictionary overlap for neural networks}
    \label{fig:venn}
\end{figure}

Figure \ref{fig:venn} shows the 4-way Venn diagram of the four models' dictionaries.
It shows that nearly there are 7,154 words in common across all models (this includes the start- and end-of-sentence tokens, as well as the unknown-word token). The advanced-filtered datasets have more unique words shared with no other models at all (866 for advanced filtering of the Py150 dataset, 942 for advanced filtering of our dataset).
Within each data set, the top 10,000 words (plus 3 special tokens) have an overlap of 7154+424+902=8480 words for the Py150 dataset, and 7154+59+820=8033 words for the new Github dataset --- i.e., advanced filtering of either dataset drops 19.7\% and 15.2\% of the vocabulary for each dataset, showing that a substantial number of common tokens occur heavily in non-linguistic contexts.

\section{Discussion}
\label{sec:discussion}
One fundamental question that can be asked about automatically generating comments is whether there is a need to distinguish the comments that are read by a human vs.\ read by a machine learning algorithm. During our work we found that source comments can have many unpredictable non-linguistic contents, and these can have noticeable impacts on language generation systems trained on this data unless these items are specifically accounted for. Great examples of this kind of items can be seen in figures \ref{fig:shortComment} and \ref{fig:manticore}. In both of these cases comments have properties that other researchers identify as useless by their algorithms \cite{qualityOfComments}: they are short, and repeat only what the code does. However, we can see from the example that in both of these cases the comments are useful, as one explains a character encoding code value in English while the other explain exactly what CPU instruction is being tested by an emulator. 

These examples on their own do not provide any linguistic utility, as neither contains full sentences, nor do they contain grammatically correct English phrases. However, even with the context of the surrounding source code the algorithmic value of these comments is questionable. If we learn that a \texttt{u'\textbackslash x8e'} character corresponds to a \texttt{CONTROL} in this case, it doesn't mean that it will be correct in a different scenario. We found that our GitHub dataset contains a total of 163 files with encoding tables and in some character encodings map this symbol to a different character. In case of the CPU instruction test example in figure \ref{fig:manticore}, the test and the comment would not be valid for a different architecture.

Using the data from our experiments we can surmise that it is, in fact, not clear in many cases if a particular comment will be useful for natural language generation or not. Source comments consisting of a proper English language without other inclusions will be very useful for machine learning algorithms (assuming it is consistent with the source code it describes). The more interesting question is whether the comments that include other data, such as formulas can eventually also be used without detriment to the resulting language generation model. 

Our hypothesis is that comments with the source and math formula inclusions are of limited use in the form they are written. However, they potentially can be transformed into a different form that can be more useful in building linguistic models. Suppose a comment contains an equation \texttt{x = 5}. It is possible to translate it into an English phrase that will fit in the context where it's used. However, this problem of translating formulas to text is still open and hard to implement, therefore further research is required and the full discussion of translating formulas to text is outside of scope of this paper.

In our experiments we have identified several problems that researchers need to be aware of if they use comments as an input to machine learning algorithms.

More than a half of the comments extracted were docstrings, the rest are single-line comments. Because Python's default parser recognizes these multiline comments simply as strings in the parse tree, extracting them requires implementing a specialized parser (which we have done). This problem is a Python-specific due to the way it standardized comments, but should not be overlooked. Having or missing this amount of data can significantly impact corpus quality.


Overall we observe from the data that comments in Python projects contain a significant amount of material that is inappropriate for use in linguistic processing.  As pointed out by \citet{dupComments}, a high number of duplicates can have up to 50\% impact on the accuracy of algorithms trained on such skewed data, when observing duplication rates of just 25\%. Our GitHub corpus contains about 40\% of duplicated comments and SriLab corpus contains about 35\% of duplicated comments. Deduplication plus our filtering removed 74\% of raw comments (advanced filtering).
Much of the duplication came from copyright notices, or editor directives such as \texttt{\# -*- coding: utf-8 -*-} (dropped due to length), as well as comments on standard ``boilerplate'' functions, such as getters, setters, methods taking user input, and methods processing and converting data. Many of these occurred both repeated verbatim, or in slight variation.

The closest point of comparison on gathering a comment corpus is the work of \citet{neuralModelGen,mcmillan:dataset}.  They prune a set of over 50 million method-comment pairs down to 2.1 million, on the basis of natural language, method size, use of Javadoc, and comment length. The described filtering process would simultaneously rule out useful data from comments not using the \lstinline[language=Java]|/**| lead for Javadoc comments and from larger methods, and would permit comments removed by the filters in Table \ref{tab:filters}.

And finally, in case many filters/steps are being applied to processing source comments, a special care needs to be taken in regards to the order in which the filtering is applied. Different order can make some filtering/processing steps useless or remove candidates that should not be removed otherwise.
\looseness=-1

\subsection{Limitations}
\label{sec:limitations}

All of the filtering patterns have the potential to discard comments containing valuable natural language information.  The code/math filters specifically target examples including punctuation or markup language, so identify comments which at a minimum contain a mixture of English and formal text. We believe this is acceptable given the difficulty of separating mixed English and formal text (see Section \ref{sec:rq4}).
Classification of natural language is inherently heuristic, given words that belong to multiple natural languages (sometimes with unrelated meanings). The character encoding check in particular runs the risk of, for example, removing English comments containing Unicode identifiers.  In our evaluation, we find relatively few comments are removed by the English filter.
\looseness=-1

In general, evaluation of the quality of a generated sentence is a difficult problem in itself. The question also has different answers depending on the context of the generation task. For example a sentence generated as a part of a story will be evaluated differently compared to a sentences that is a part of a source code comment. The quality of the source code comments is a widely discussed topic and there are many criteria used, such as relevance of the comment to the source it describes. In our case we are only interested in evaluating the contents of the source comments on their own as English sentences. The widely-used metric for this kind of evaluation is BLEU score, but it only evaluates if generated sentence matches the "golden standard" which might or might not be a high-quality sentence itself. BLEU scores have some issues and can be misleading if used by themselves ~\cite{callison2006re,callison2008further,sulem2018bleu, post2018call}. We also used parsing as a means to check the level of grammaticallity of the sentences, however grammatical sentence might not be evaluated as a high quality. As an example, a short grammatical sentence might be viewed as too simplistic. Overall the definition of a "high-quality" sentence is a little blurred and the current latest work evaluates generated sentences by their syntactic or semantic similarity to the original. Our goal was evaluation of the general quality of a sentence and thus most of the current evaluation methods are not really suited and we are limited in our analysis by the technology.

Our evaluation considers the largest bodies of general source code comments we know of,
which is a double-edged sword.  On one hand, its size and scope balances out the fact that prior work has evaluated on comment data that is both limited in size and biased towards high-quality (for human use) comments --- these are the reasons we chose the corpora we used.
On the other hand, selecting the 840 most popular Python projects on Github naturally selects for other biases.  The projects studied are open source; as usual for such cases, proprietary software may have different trends. The projects were selected based on popularity, not any measure of quality, which is generally unknown, so the corpus may include pathological cases well outside the minimum level of quality where we might reasonably expect comment processing or generating tools to function well.  This is balanced out to a degree by also evaluating on the SRILab Py150 dataset, which is smaller, but also sizeable and selected on very different criteria.

\citet{neuralModelGen,mcmillan:dataset} make a point of removing auto-generated code from their corpus by removing any file containing ``generated by'', following \citet{shimonaka2016identifying}.  This filter is extremely coarse. There are 20,001 occurrence of this string in the 19,268 Python files in our GitHub dataset, of which a large number are meaningful comments (e.g., comments discussing data generated by other systems or the method by which a shape is being generated). For those clearly indicating generated source code, the mentioned tools do not appear to add significant numbers of comments to source code, so we left these in our data.  However, this may not be a generalizable trend, and may in fact be specific to the tools used to generate Python (vs. Java) in our dataset. At the same time, we observe some auto-generated code (e.g., from ANTLR) which lacks this phrase, so this was already an imprecise filter.
\looseness=-1

Our process for answering Research Question 1 (non-linguistic categories) relies on manual analysis of a randomly selected subset of the corpus, which naturally admits the possibility that other interesting categories of non-linguistic comments may have been missed. Likewise, our process for Research Question 2 (filters for those categories) also relies on manual inspection of the filters' effects on the subset from Research Question 1 and heuristic manual searches through the remainder of the corpus, which may have missed comments which were incorrectly filtered (or not filtered).

Our evaluation of Research Question 5 considers two kinds of techniques, but only one goal task: comment completion.  Evaluation on a different kind of task (e.g., oracle synthesis, or generating a comment from code) may yield different trends (e.g., a certain kind of filtering becoming more or less important), or require new task-specific filters (e.g., for some approximation of language quality).  However, these would also require a slightly different sort of corpus, such as one pairing code with comments~\cite{mcmillan:dataset}. We believe conclusions from the comment completion tasks likely generalize to other tasks involving comments, even if they require other inputs as well.

\section{Conclusions}
For building tools consuming or producing natural language from comments, the quality of the comments used for training or evaluation can be a significant factor. We found that there are a number of substantial categories of comments which may be useful in some ways, but are not purely linguistic in nature, making them undesirable in linguistically-oriented datasets as they are written.  We offer a categorization of non-obvious linguistically-undesirable comments, and practical filters for removing them. Finally, we have shown that applying these filters to remove non-linguistic data improves the output quality of two generative models (4-gram and neural network models), on two large datasets of open source Python comments.  We believe this offers compelling evidence that software engineering tools trained or evaluated on source code comments should be closer attention to the linguistic qualities of the data than is typically done today, and have provided a set of reusable filters to remove non-linguistic data to help.


\bibliographystyle{ACM-Reference-Format}
\bibliography{references}

\end{document}